\documentclass[review]{elsarticle}

\usepackage{lineno,hyperref}

\usepackage{amsmath,amsthm,amsfonts,bm}	 
\usepackage{subfigure}
\usepackage{float}
\usepackage{multirow}

\journal{International Journal of Heat and Mass Transfer}

\begin{document}

\begin{frontmatter}

\title{Lattice Boltzmann simulations of three-dimensional thermal convective flows at high Rayleigh number
\footnote{\href{https://doi.org/10.1016/j.ijheatmasstransfer.2019.06.002}{DOI: 10.1016/j.ijheatmasstransfer.2019.06.002}}
\footnote{\copyright \ 2019. This manuscript version is made available under the CC-BY-NC-ND 4.0 license http://creativecommons.org/licenses/by-nc-nd/4.0/}}



\author[mymainaddress]{Ao Xu\corref{mycorrespondingauthor}}
\cortext[mycorrespondingauthor]{Corresponding author}
\ead{axu@nwpu.edu.cn}

\author[mysecondaryaddress]{Le Shi}

\author[mymainaddress]{Heng-Dong Xi}

\address[mymainaddress]{School of Aeronautics, Northwestern Polytechnical University, Xi'an 710072, China}
\address[mysecondaryaddress]{State Key Laboratory of Electrical Insulation and Power Equipment, Center of Nanomaterials for Renewable Energy, School of Electrical Engineering, Xi'an Jiaotong University, Xi'an 710049, China}

\begin{abstract}
We present numerical simulations of three-dimensional thermal convective flows in a cubic cell at high Rayleigh number using thermal lattice Boltzmann (LB) method.
The thermal LB model is based on double distribution function approach, which consists of a D3Q19 model for the Navier-Stokes equations to simulate fluid flows and a D3Q7 model for the convection-diffusion equation to simulate heat transfer.
Relaxation parameters are adjusted to achieve the isotropy of the fourth-order error term in the thermal LB model.
Two types of thermal convective flows are considered: one is laminar thermal convection in side-heated convection cell, which is heated from one vertical side and cooled from the other vertical side; while the other is turbulent thermal convection in Rayleigh-B\'enard convection cell, which is heated from the bottom and cooled from the top.
In side-heated convection cell, steady results of hydrodynamic quantities and Nusselt numbers are presented at Rayleigh numbers of $10^{6}$ and $10^{7}$, and Prandtl number of 0.71, where the mesh sizes are up to $257^{3}$;
in Rayleigh-B\'enard convection cell, statistical averaged results of Reynolds and Nusselt numbers, as well as kinetic and thermal energy dissipation rates are presented at Rayleigh numbers of $10^{6}$, $3\times 10^{6}$, and $10^{7}$, and Prandtl numbers of 0.7 and 7, where the nodes within thermal boundary layer are around 8.
Compared with existing benchmark data obtained by other methods, the present LB model can give consistent results.
\end{abstract}

\begin{keyword}
Lattice Boltzmann method \sep Thermal convective flows \sep Three dimension \sep High Rayleigh number
\end{keyword}

\end{frontmatter}


\section{Introduction}

Thermal convective flows occur ubiquitously in nature and has wide applications in industry \cite{ahlers2009heat,lohse2010small}.
An in-depth understanding of the complex transport mechanism in thermal convective flows requires powerful experimental and computational tools.
Over the past three decades, the lattice Boltzmann (LB) method has attracted broad interest in computational fluid dynamics and numerical heat transfer communities due to its ability to simulate complex flows, as well as easy implementation on various parallel programming platforms \cite{chen1998lattice,aidun2010lattice,xu2017lattice,cheng2014recent}.

The early effort to construct LB model for thermal convective flows focused on energy-conserving LB models, where fluid density, velocity, and temperature are obtained from various moments of the distribution function $f_{i}$ \cite{alexander1993lattice,chen1994thermal}.
Compared with the LB model for isothermal flows, a larger set of discrete velocities was adopted to obtain the evolution equation of temperature.
However, due to the spurious coupling between shear and energy modes, it was observed that the energy-conserving LB models suffer severe numerical instability issue \cite{lallemand2003theory}.
To avoid this issue, an alternative approach is to treat the temperature as a scalar when the viscous heat dissipation and compression work done by the pressure are negligible.
As a result, the temperature field is governed by a convection diffusion equation (CDE), and one may either use a hybrid scheme or a double distribution function (DDF) scheme to obtain the temperature.
In both schemes, conventional isothermal LB model is adopted to solve fluid flows, which is essentially governed by the Navier-Stokes equations at macroscopic level.
The difference between the hybrid scheme and the DDF scheme is that, the finite difference (FD) method is adopted to solve the target temperature CDE in the hybrid scheme, while an additional distribution function for the temperature field is introduced in the DDF scheme.
In the LB-FD hybrid scheme, implementing temperature boundary condition is nontrivial, since boundary nodes will not overlap for flow and temperature fields.
Specifically, the FD method requires implementing temperature boundary condition at the fluid-solid interface, while the LB model adopts popular half-way bounce-back scheme to mimic no-slip velocity boundary and it requires implementing velocity boundary condition half-lattice off the fluid-solid interface \cite{he1997analytic}.

For the above reasons, the DDF scheme based LB models have been widely adopted to simulate thermal convective flows.
The early work of Shan \cite{shan1997simulation} employed a two-component LB model where one component represents the motion of the fluid and the other component simulates a passive temperature field.
Guo et al. \cite{guo2002coupled} constructed a thermal LB model based on incompressible LB model to reduce compressibility errors.
Through Chapman-Enskog analysis, the incompressible Navier-Stokes equations under the Boussinesq assumption as well as the CDE for temperature can be obtained.
Recently, Chai and Zhao \cite{chai2013lattice} modified equilibrium distribution function and used an additional source term to recover the CDE.
Huang and Wu \cite{huang2015lattice} proposed to remove the deviation term in the corresponding macroscopic CDE via treating the divergence-free velocity field as a source term in the LB equation.
In addition to isotropic diffusion problems, efforts have been taken to solve anisotropic CDEs via adopting the two-relaxation-time (TRT) collision operator (e.g., the previous work of Ginzburg \cite{ginzburg2005equilibrium}) and multiple-relaxation-time (MRT) collision operator (e.g., the previous work of Rasin et al. \cite{rasin2005multi}, Yoshida and Nagaoka \cite{yoshida2010multiple}, Huang and Wu \cite{huang2014modified}).
By adjusting the relaxation rates in the MRT relaxation matrix, isotropy for the fourth-order error term in corresponding macroscopic CDE can be attained \cite{dubois2009towards,ginzburg2010optimal}.
Wang et al. \cite{wang2013lattice} simulated the incompressible thermal flows in two-dimensional (2D) square cavity under the Boussinesq approximation. Contrino et al. \cite{contrino2014lattice} then used the same approach to simulate thermally driven 2D square cavity at high Rayleigh number, and they provided results of benchmark qualities.

In this work, we proposed a three-dimensional (3D) double distribution function (DDF) based LB model to simulate thermal convective flows.
A D3Q19 model for the Navier-Stokes equations to simulate fluid flows and a D3Q7 model for the convection-diffusion equation to simulate heat transfer were adopted.
To ensure the stability of the numerical model, relaxation parameters were adjusted to enforce fourth-order accuracy of the thermal model \cite{dubois2009towards,ginzburg2010optimal}.
With this thermal LB model, we simulated the following two types of thermal flows:
one is thermal flows in a cubic cell with differentially heated vertical walls, the other is Rayleigh-B\'enard convection in a cubic cell heated from the below and cooled from the above.
It should be noted that both flow configurations have been adopted as canonic flow systems for studying thermal flows.
Here we aim to provide benchmark quality results for thermal convective flows in the cubic cell.
The rest of the paper is organized as follows: In Section 2, we first present the 3D double-distribution multiple-relaxation-time LB model for simulating fluid flows and heat transfer.
In Section 3, laminar thermal convection in side-heated convection cell at Rayleigh numbers of Ra = $10^{6}$ and $10^{7}$, and Prandtl number of Pr = 0.71 are simulated.
The convergence behavior of steady results is obtained with grid up to $257^{3}$.
In Section 4, turbulent thermal convection in Rayleigh-B\'enard convection cell are simulated at Ra = $10^{6}$, $3\times 10^{6}$ and $10^{7}$; Pr is fixed as 0.7 and 7, which corresponds to the working fluids of air and water at $20^{\circ}$C, respectively.
The statistically averaged flow and temperature quantities, as well as energy dissipation rates are provided.

\section{Numerical method}

\subsection{Mathematical model for incompressible thermal flows}

In incompressible thermal flows, temperature variation will cause density variation, thus resulting in buoyancy effect.
Following the Boussinesq approximation, the temperature can be treated as an active scalar and its influence to the velocity field is realized through the buoyancy term.
The viscous heat dissipation and compression work due to pressure are therefore neglected.
All the transport coefficients are assumed to be constants. Then, the governing equations can be written as
\begin{subequations}
\begin{align}
& \nabla \cdot \mathbf{u}=0 \\
& \frac{\partial \mathbf{u}}{\partial t}+\mathbf{u}\cdot \nabla \mathbf{u}=-\frac{1}{\rho_{0}}\nabla p+\nu \nabla^{2}\mathbf{u}+g\beta_{T}(T-T_{0})\hat{\mathbf{z}} \\
& \frac{\partial T}{\partial t}+\mathbf{u}\cdot \nabla T=\kappa \nabla^{2} T
\end{align} \label{Eq.NS}
\end{subequations}
where $\mathbf{u}$, $p$, and $T$ are the fluid velocity, pressure and temperature, respectively.
$\rho_{0}$ and $T_{0}$  are reference density and temperature, respectively.
$\nu$, $\beta_{T}$ and $\kappa$ are the kinematic viscosity, thermal expansion coefficient and thermal diffusivity, respectively.
$g$ is the gravity value, and $\hat{\mathbf{z}}$  is unit vector in the vertical direction.

With the scalings
\begin{equation}
    \begin{split}
&  \mathbf{x}/L_{0} \rightarrow \mathbf{x}^{*}, \ \ t/\sqrt{L_{0}/(g\beta_{T}\Delta_{T})} \rightarrow t^{*}, \ \ \mathbf{u}/\sqrt{g\beta_{T}L_{0}\Delta_{T}} \rightarrow \mathbf{u}^{*}, \\
& p/(\rho_{0}g\beta_{T}\Delta_{T}L_{0}) \rightarrow p^{*}, \ \ (T-T_{0})/\Delta_{T} \rightarrow T^{*}
    \end{split}
\end{equation}
Then, Eq. \ref{Eq.NS} can be rewritten in dimensionless form as
\begin{subequations}
\begin{align}
& \nabla \cdot \mathbf{u}^{*}=0 \\
& \frac{\partial \mathbf{u}^{*}}{\partial t}+\mathbf{u}^{*}\cdot \nabla \mathbf{u}^{*}=-\nabla p^{*}+\sqrt{\frac{\text{Pr}}{\text{Ra}}} \nabla^{2}\mathbf{u}^{*}+T^{*}\tilde{\mathbf{z}} \\
& \frac{\partial T^{*}}{\partial t}+\mathbf{u}^{*}\cdot \nabla T^{*}=\sqrt{\frac{1}{\text{Pr}\text{Ra}}} \nabla^{2} T
\end{align}
\end{subequations}
where the dimensionless numbers characterizing the system are Rayleigh and Prandtl numbers, defined as
\begin{equation}
\text{Ra}=\frac{g\beta_{T} \Delta_{T}L^{3}_{0}}{\nu \kappa}, \ \ \text{Pr}=\frac{\nu}{\kappa}
\end{equation}

\subsection{The LB model for fluid flows}
In the LB method, to solve Eqs. \ref{Eq.NS}a and \ref{Eq.NS}b, the evolution equation of density distribution function is written as
\begin{equation}
  f_{i}(\mathbf{x}+\mathbf{e}_{i}\delta_{t},t+\delta_{t})-f_{i}(\mathbf{x},t)=-(\mathbf{M}^{-1}\mathbf{S})_{ij}\left[\mathbf{m}_{j}(\mathbf{x},t)-\mathbf{m}_{j}^{(\text{eq})}(\mathbf{x},t)\right]
  +\delta_{t}F_{i}^{'} \label{Eq.MRT}
\end{equation}
where $f_{i}$ is the density distribution function.
$\mathbf{x}$ is the fluid parcel position, $t$ is the time, $\delta_{t}$ is the time step.
$\mathbf{e}_{i}$ is the discrete velocity along the $i$th direction.
For the three-dimensional D3Q19 discrete velocity model, $\mathbf{e}_{i}$ can be given as
\begin{equation}
\setlength{\arraycolsep}{3pt}
\setcounter{MaxMatrixCols}{20}
\begin{split}
&
\big[\mathbf{e}_{0},\mathbf{e}_{1},\mathbf{e}_{2},\mathbf{e}_{3},\mathbf{e}_{4},\mathbf{e}_{5},\mathbf{e}_{6},\mathbf{e}_{7},\mathbf{e}_{8},\mathbf{e}_{9},\mathbf{e}_{10},
\mathbf{e}_{11},\mathbf{e}_{12},\mathbf{e}_{13},\mathbf{e}_{14},\mathbf{e}_{15},\mathbf{e}_{16},\mathbf{e}_{17},\mathbf{e}_{18} \big]=
\\
&c
\begin{bmatrix}
0 & 1 & -1 &  0 &  0 &  0 &  0 &  1 & -1 &  1 & -1 & 1 & -1 &  1 & -1 &  0 &  0 &  0 &  0 \\
0 & 0 &  0 &  1 & -1 &  0 &  0 &  1 &  1 & -1 & -1 & 0 &  0 &  0 &  0 &  1 & -1 &  1 & -1 \\
0 & 0 &  0 &  0 &  0 &  1 & -1 &  0 &  0 &  0 &  0 & 1 &  1 & -1 & -1 &  1 &  1 & -1 & -1 \\
\end{bmatrix}
\end{split}
\end{equation}
In the above, $c=\delta_{x}/\delta_{t}$ is the lattice constant.
For simplicity, we adopt $c=\delta_{x}=\delta_{t}=1$.
$\mathbf{M}$ is a $19 \times 19$ orthogonal transformation matrix, and it is given by
\begin{small}
\begin{equation}
\setlength{\arraycolsep}{2.0pt}
\setcounter{MaxMatrixCols}{19}
\centering
\begin{split}
&\mathbf{M}
= \bigg[
 \langle 1 |, \ \langle 19\mathbf{e}^{2}-30 |, \ \langle \frac{21}{2}\mathbf{e}^{4}-\frac{53}{2}\mathbf{e}^{2}+12 |, \
\langle \mathbf{e}_{x} |, \ \langle (5\mathbf{e}^{2}-9)\mathbf{e}_{x} |, \
\langle \mathbf{e}_{y} |, \ \langle (5\mathbf{e}^{2}-9)\mathbf{e}_{y} |, \ \langle \mathbf{e}_{z} |,  \\
& \ \ \ \ \ \ \ \ \
\langle (5\mathbf{e}^{2}-9)\mathbf{e}_{z} |, \
\langle 3\mathbf{e}_{x}^{2}-\mathbf{e}^{2} |, \ \langle (3\mathbf{e}^{2}-5)(3\mathbf{e}_{x}^{2}-\mathbf{e}^{2}) |, \
\langle \mathbf{e}_{y}^{2}-\mathbf{e}_{z}^{2} |, \ \langle (3\mathbf{e}^{2}-5)(\mathbf{e}_{y}^{2}-\mathbf{e}_{z}^{2}) |, \\
& \ \ \ \ \ \ \ \ \
\langle \mathbf{e}_{x}\mathbf{e}_{y} |, \ \langle \mathbf{e}_{y}\mathbf{e}_{z} |, \ \langle \mathbf{e}_{x}\mathbf{e}_{z} |, \
\langle (\mathbf{e}_{y}^{2}-\mathbf{e}_{z}^{2})\mathbf{e}_{x} |, \
\langle (\mathbf{e}_{z}^{2}-\mathbf{e}_{x}^{2})\mathbf{e}_{y} |, \
\langle (\mathbf{e}_{x}^{2}-\mathbf{e}_{y}^{2})\mathbf{e}_{z} | \bigg]^{T} = \\
&    \begin{pmatrix}
    1   &  1  &  1  &  1  &  1  &  1  &  1  &  1 &  1  &  1 &   1 &  1 &   1 &  1 &   1 & 1  & 1  & 1 &  1 \\ 
    -30 & -11 & -11 & -11 & -11 & -11 & -11 &  8 &  8  &  8 &   8 &  8 &   8 &  8 &   8 & 8  & 8  & 8 &  8 \\ 
    12  & -4  & -4  & -4  & -4  & -4  & -4  &  1 &  1  &  1 &   1 &  1 &   1 &  1 &   1 & 1  & 1  & 1 &  1 \\ 
    0   &  1  & -1  &  0  &  0  &  0  &  0  &  1 &  -1 &  1 &  -1 &  1 &  -1 &  1 &  -1 & 0  & 0  & 0 &  0 \\ 
    0   & -4  &  4  &  0  &  0  &  0  &  0  &  1 &  -1 &  1 &  -1 &  1 &  -1 &  1 &  -1 & 0  & 0  & 0 &  0 \\ 
    0   &  0  &  0  &  1  & -1  &  0  &  0  &  1 &   1 & -1 &  -1 &  0 &  0  &  0 &   0 & 1  & -1 & 1 & -1 \\ 
    0   &  0  &  0  & -4  &  4  &  0  &  0  &  1 &   1 & -1 &  -1 &  0 &  0  &  0 &   0 & 1  & -1 & 1 & -1 \\ 
    0   &  0  &  0  &  0  &  0  &  1  &  -1 &  0 &   0 &  0 &   0 &  1 &   1 & -1 &  -1 & 1  & 1  & -1 & -1 \\ 
    0   &  0  &  0  &  0  &  0  & -4  &  4  &  0 &   0 &  0 &   0 &  1 &   1 & -1 &  -1 & 1  & 1  & -1 & -1 \\ 
    0   &  2  &  2  & -1  & -1  & -1  & -1  &  1 &   1 &  1 &   1 &  1 &   1 &  1 &   1 & -2 & -2 & -2 & -2 \\ 
    0   & -4  & -4  &  2  &  2  &  2  &  2  &  1 &   1 &  1 &   1 &  1 &   1 &  1 &   1 & -2 & -2 & -2 & -2 \\ 
    0   &  0  &  0  &  1  &  1  & -1  & -1  &  1 &   1 &  1 &   1 & -1 &  -1 & -1 &  -1 &  0 &  0 & 0  & 0  \\ 
    0   &  0  &  0  & -2  & -2  &  2  &  2  &  1 &   1 &  1 &   1 & -1 &  -1 & -1 &  -1 &  0 &  0 & 0  & 0  \\ 
    0   &  0  &  0  &  0  &  0  &  0  &  0  &  1 &  -1 & -1 &   1 &  0 &   0 &  0 &   0 &  0 &  0 &  0  & 0  \\ 
    0   &  0  &  0  &  0  &  0  &  0  &  0  &  0 &   0 &  0 &   0 &  0 &   0 &  0 &   0 &  1 & -1 & -1  & 1  \\ 
    0   &  0  &  0  &  0  &  0  &  0  &  0  &  0 &   0 &  0 &   0 &  1 &  -1 & -1 &   1 &  0 &  0 &  0  & 0  \\ 
    0   &  0  &  0  &  0  &  0  &  0  &  0  &  1 &  -1 &  1 &  -1 & -1 &   1 & -1 &   1 &  0 &  0  &  0  & 0  \\ 
    0   &  0  &  0  &  0  &  0  &  0  &  0  & -1 &  -1 &  1 &   1 &  0 &   0 &  0 &   0 &  1 &  -1 &  1  & -1 \\ 
    0   &  0  &  0  &  0  &  0  &  0  &  0  &  0 &   0 &  0 &   0 &  1 &   1 & -1 &  -1 & -1 &  -1 &  1  &  1 \\ 
    \end{pmatrix}
\end{split}
\end{equation}
\end{small}
Choose the equilibrium distribution function as $f_{i}^{(\text{eq})}=\omega_{i}\rho\left[1+\frac{\mathbf{e}_{i}\cdot\mathbf{u}}{c_{s}^{2}}
+\frac{(\mathbf{e}_{i}\cdot\mathbf{u})^{2}}{2c_{s}^{4}}
-\frac{|\mathbf{u}|^{2}}{2c_{s}^{2}} \right]$, where the weights are $\omega_{0}=1/3$, $\omega_{1-6}=1/18$, $\omega_{7-18}=1/36$.
Then, the equilibrium moments $\mathbf{m}^{(\text{eq})}$ are
\begin{equation}
    \begin{split}
\mathbf{m}^{(\text{eq})}
=\rho \bigg[
& 1,\ -11+19|\mathbf{u}|^{2},\ 3-\frac{11}{2}|\mathbf{u}|^{2},\ u,\ -\frac{2}{3}u,\ v,\ -\frac{2}{3}v, \ w,\\
& -\frac{2}{3}w, \ 2u^{2}-v^{2}-w^{2},\ -\frac{1}{2}(2u^{2}-v^{2}-w^{2}), \ v^{2}-w^{2}, \\
& -\frac{1}{2}(v^{2}-w^{2}), \ uv, \ vw, \ uw, \ 0,\ 0, \ 0 \bigg]^{T} \\
    \end{split}
\end{equation}
The diagonal relaxation matrix $\mathbf{S}$ is given as
\begin{equation}
\mathbf{S}=\text{diag}(s_{\rho},s_{e},s_{\varepsilon},s_{j},s_{q},s_{j},s_{q},s_{j},s_{q},s_{\nu},s_{\pi},s_{\nu},s_{\pi},s_{\nu},s_{\nu},s_{\nu},s_{m},s_{m},s_{m})
\end{equation}
To ensure accurate flow boundary conditions as well as adequate numerical stability, relaxation parameters $s_{i}$
are choosen as $s_{\rho}=s_{j}=0$, $s_{e}=s_{\varepsilon}=s_{\nu}=s_{\pi}=1/\tau_{f}$, $s_{q}=s_{m}=8(2\tau_{f}-1)/(8\tau_{f}-1)$.
Here, $\tau_{f}$ is determined by the kinematic viscosity of the fluids as $\nu=c_{s}^{2}(\tau_{f}-0.5)\delta_{t}$, and $c_{s}=1/\sqrt{3}c$ is the speed of sound.
The forcing term $F_{i}^{'}$ in the right-hand side of Eq. \ref{Eq.MRT} is given by
\begin{equation}
\mathbf{F}^{'}=\mathbf{M}^{-1}\left( \mathbf{I}-\frac{\mathbf{S}}{2} \right)\mathbf{M}\tilde{\mathbf{F}}
\end{equation}
and the term $\mathbf{M\tilde{F}}$ is \cite{guo2002discrete,guo2008analysis}
\begin{equation}
    \begin{split}
\mathbf{M}\bar{\mathbf{F}}=\bigg[
& 0,\   38\mathbf{u}\cdot\mathbf{F},\  -11\mathbf{u}\cdot\mathbf{F},\  F_{x},\  -\frac{2}{3}F_{x},\  F_{y},\  -\frac{2}{3}F_{y},  \ F_{z}, \ -\frac{2}{3}F_{z}, \\
& \ 4uF_{x}-2vF_{y}-2wF_{z},\  -2uF_{x}+vF_{y}+wF_{z},  \ 2vF_{y}-2wF_{z}, \\
& \ -vF_{y}+wF_{z},  \ uF_{y}+vF_{x},  \ vF_{z}+wF_{y},  \ uF_{z}+wF_{x},  \ 0, \ 0, \ 0  \bigg]^{T}
    \end{split}
\end{equation}
where $\mathbf{F}=\rho g\beta_{T}(T-T_{0})\hat{\mathbf{z}}$.
The macroscopic density $\rho$ and velocity $\mathbf{u}$ are obtained from
\begin{equation}
  \rho=\sum_{i=0}^{18}f_{i}, \ \ \mathbf{u}=\frac{1}{\rho}\left( \sum_{i=0}^{18}\mathbf{e}_{i}f_{i}+\frac{1}{2}\mathbf{F} \right)
\end{equation}

The no-slip velocity boundary conditions at the wall can be realized by the half-way bounce-back boundary scheme as
\begin{equation}
  f_{\bar{i}}(\mathbf{x}_{f},t+\delta_{t})=f_{i}^{+}(\mathbf{x}_{f},t)
\end{equation}
where $f_{i}^{+}(\mathbf{x}_{f},t)$ is the post collision value of the distribution function, $f_{\bar{i}}(\mathbf{x}_{f},t)$ is the distribution function associated with the velocity $\mathbf{e}_{\bar{i}}=-\mathbf{e}_{i}$.

\subsection{The LB model for heat transfer}
To solve Eq. \ref{Eq.NS}c, the evolution equation of temperature distribution function is written as
\begin{equation}
  g_{i}(\mathbf{x}+\mathbf{e}_{i}\delta_{t},t+\delta_{t})-g_{i}(\mathbf{x},t)=-(\mathbf{N}^{-1}\mathbf{Q})_{ij}\left[\mathbf{n}_{j}(\mathbf{x},t)-\mathbf{n}_{j}^{(\text{eq})}(\mathbf{x},t)\right]
  \label{Eq.MRT_T}
\end{equation}
where $g_{i}$ is the temperature distribution function.
For the three-dimensional D3Q7 discrete velocity model, $\mathbf{e}_{i}$ can be given as
\begin{equation}
\setlength{\arraycolsep}{3pt}
\setcounter{MaxMatrixCols}{20}
\big[\mathbf{e}_{0},\mathbf{e}_{1},\mathbf{e}_{2},\mathbf{e}_{3},\mathbf{e}_{4},\mathbf{e}_{5},\mathbf{e}_{6} \big]=
c
\begin{bmatrix}
0 & 1 & -1 &  0 &  0 &  0 &  0  \\
0 & 0 &  0 &  1 & -1 &  0 &  0  \\
0 & 0 &  0 &  0 &  0 &  1 & -1  \\
\end{bmatrix}
\end{equation}
$\mathbf{N}$ is a $7 \times 7$ orthogonal transformation matrix, and it is given by
\begin{equation}
\setlength{\arraycolsep}{2.0pt}
\setcounter{MaxMatrixCols}{19}
\centering
\mathbf{N}
 = \begin{pmatrix}
 \langle 1 | \\
 \langle \mathbf{e}_{x} | \\
 \langle \mathbf{e}_{y} | \\
 \langle \mathbf{e}_{z} | \\
 \langle -6+7\mathbf{e}^{2} | \\
 \langle 3 \mathbf{e}_{x}^{2}-\mathbf{e}^{2} | \\
 \langle \mathbf{e}_{y}^{2}-\mathbf{e}_{z}^{2} |
    \end{pmatrix}
 =   \begin{pmatrix}
    1   &  1  &  1  &  1  &  1  &  1  &  1   \\ 
    0   &  1  & -1  &  0  &  0  &  0  &  0   \\ 
    0   &  0  &  0  &  1  & -1  &  0  &  0   \\ 
    0   &  0  &  0  &  0  &  0  & -1  & -1   \\ 
   -6   &  1  &  1  &  1  &  1  &  1  &  1   \\ 
    0   &  2  &  2  & -1  & -1  & -1  & -1   \\ 
    0   &  0  &  0  &  1  &  1  & -1  & -1   \\ 
    \end{pmatrix}
\end{equation}
Choose the equilibrium distribution function as $g_{i}^{(\text{eq})}=\omega_{i}T\left[1+\frac{7}{6+a_{T}}\frac{\mathbf{e}_{i}\cdot\mathbf{u}}{c_{s}^{2}} \right]$, where the weights are $\omega_{0}=(1-a_{T})/7$, $\omega_{1-6}=(6+a_{T})/42$.
Then, the equilibrium moments $\mathbf{n}^{(\text{eq})}$ are
\begin{equation}
\mathbf{n}^{(\text{eq})}
=\left[ T, \ uT, \ vT, \ wT, \ a_{T}T, \ 0, \ 0  \right]^{T}
\end{equation}
where $a_{T}$ is a constant.
The relaxation matrix is given by $\mathbf{Q}=\text{diag}(0,q_{\kappa},q_{\kappa},q_{\kappa},q_{e},q_{\nu},q_{\nu})$.
The thermal diffusivity $\kappa$ is determined from the relaxation parameter $q_{\kappa}$ as
\begin{equation}
  \kappa=\frac{6+a_{T}}{21}\left(\frac{1}{q_{\kappa}}-\frac{1}{2} \right)
\end{equation}
To achieve the isotropy of the fourth-order error term, Dubois et al. \cite{dubois2009towards} proposed the following relationships for the relaxation parameters in D3Q7 model:
\begin{equation}
\left( \frac{1}{q_{\kappa}}-\frac{1}{2} \right)\left( \frac{1}{q_{e}}-\frac{1}{2} \right)=\frac{1}{6} \label{Eq.magic}
\end{equation}
\begin{equation}
\frac{1}{q_{\nu}}-\frac{1}{2}=\frac{a_{T}+6}{1-a_{T}}\left( \frac{1}{q_{\kappa}}-\frac{1}{2} \right)-\frac{4+3a_{T}}{12(1-a_{T})}\left( \frac{1}{q_{\kappa}}-\frac{1}{2} \right)^{-1} \label{Eq.4th}
\end{equation}
From Eq. \ref{Eq.4th}, we have
\begin{equation}
  q_{\nu} = \frac{6(1-a_{T})(2-q_{\kappa})q_{\kappa}}{(11+3a_{T})(q_{\kappa}-6)q_{\kappa}+12(a_{T}+6)} \label{Eq.qk}
\end{equation}
If and only if we take a special value of $q_{\kappa}$ as
\begin{equation}
  \frac{1}{q_{\kappa}}-\frac{1}{2}=\frac{\sqrt{3}}{6}
\end{equation}
then $q_{\nu}$ in Eq. \ref{Eq.qk} becomes a constant independent of $a_{T}$, which is
\begin{equation}
  \frac{1}{q_{\nu}}-\frac{1}{2}=\frac{\sqrt{3}}{3} \label{Eq.qnu}
\end{equation}
With Eq. \ref{Eq.qnu}, we can determine $q_{e}$ from Eq. \ref{Eq.magic} as
\begin{equation}
  \frac{1}{q_{e}}-\frac{1}{2}=\frac{\sqrt{3}}{3}
\end{equation}
In short, we have $q_{\kappa}=3-\sqrt{3}$, $q_{e}=q_{\nu}=4\sqrt{3}-6$ and $a_{T}=42\sqrt{3}\kappa-6$.
The macroscopic temperature $T$ is obtained from
\begin{equation}
  T=\sum_{i=0}^{6}g_{i}
\end{equation}

The Dirichlet boundary conditions for constant temperature can be realized by the half-way anti-bounce-back boundary scheme as \cite{li2013boundary}
\begin{equation}
  g_{\bar{i}}(\mathbf{x}_{f},t+\delta_{t})=-g_{i}^{+}(\mathbf{x}_{f},t)+\frac{6+a_{T}}{21}T_{w}
\end{equation}
where $T_{w}$ is the wall temperature.
The Neumann boundary conditions for adiabatic temperature can be realized by the half-way bounce-back scheme as
\begin{equation}
  g_{\bar{i}}(\mathbf{x}_{f},t+\delta_{t})=g_{i}^{+}(\mathbf{x}_{f},t)
\end{equation}

\section{Laminar thermal convection in side-heated convection cell}

The flow configuration for the side-heated convection cell is shown in Fig. 1.
The left and right vertical walls are kept at constant hot and cold temperature, respectively; the other four walls are adiabatic.
All six walls impose no-slip velocity boundary condition. The dimension of the cell is $L \times D \times H$, and we set $L=D=H$ in this work.
Simulation results are provided at Rayleigh numbers of $10^{6}$ and $10^{7}$; the Prandtl number is fixed as 0.71.
In addition, we need another dimensionless parameter, the Mach number that is defined as  $\text{Ma}=\sqrt{g\beta_{T}L_{0}\Delta_{T}}/c_{s}$, to fully determine the parameters in the system.
Here, we fix Ma = 0.1 as a compromise to approximate the incompressibility condition as well as to enhance the computational efficiency.
With the above parameters in side-heated convection cell, steady state can be achieved for which the criterion is given as
\begin{equation}
    \begin{split}
& \frac{\sum_{i}\left\|\mathbf{u}(\mathbf{x}_{i},t+2000\delta_{t})-\mathbf{u}(\mathbf{x}_{i},t)\right\|_{2}}{\sum_{i}\left\|\mathbf{u}(\mathbf{x}_{i},t)\right\|_{2}}<10^{-9}, \\
& \frac{\sum_{i}|T(\mathbf{x}_{i},t+2000\delta_{t})-T(\mathbf{x}_{i},t)|_{1}}{\sum_{i}|T(\mathbf{x}_{i},t)|_{1}}<10^{-9} \\
    \end{split}
\end{equation}
where $\left\| \mathbf{u} \right\|_{2}$ denotes $L^{2}$ norm of $\mathbf{u}$,
and $\left| T \right|_{1}$ denotes $L^{1}$ norm of $T$.
\begin{figure}[H]
  \centering
  \includegraphics[width=6cm]{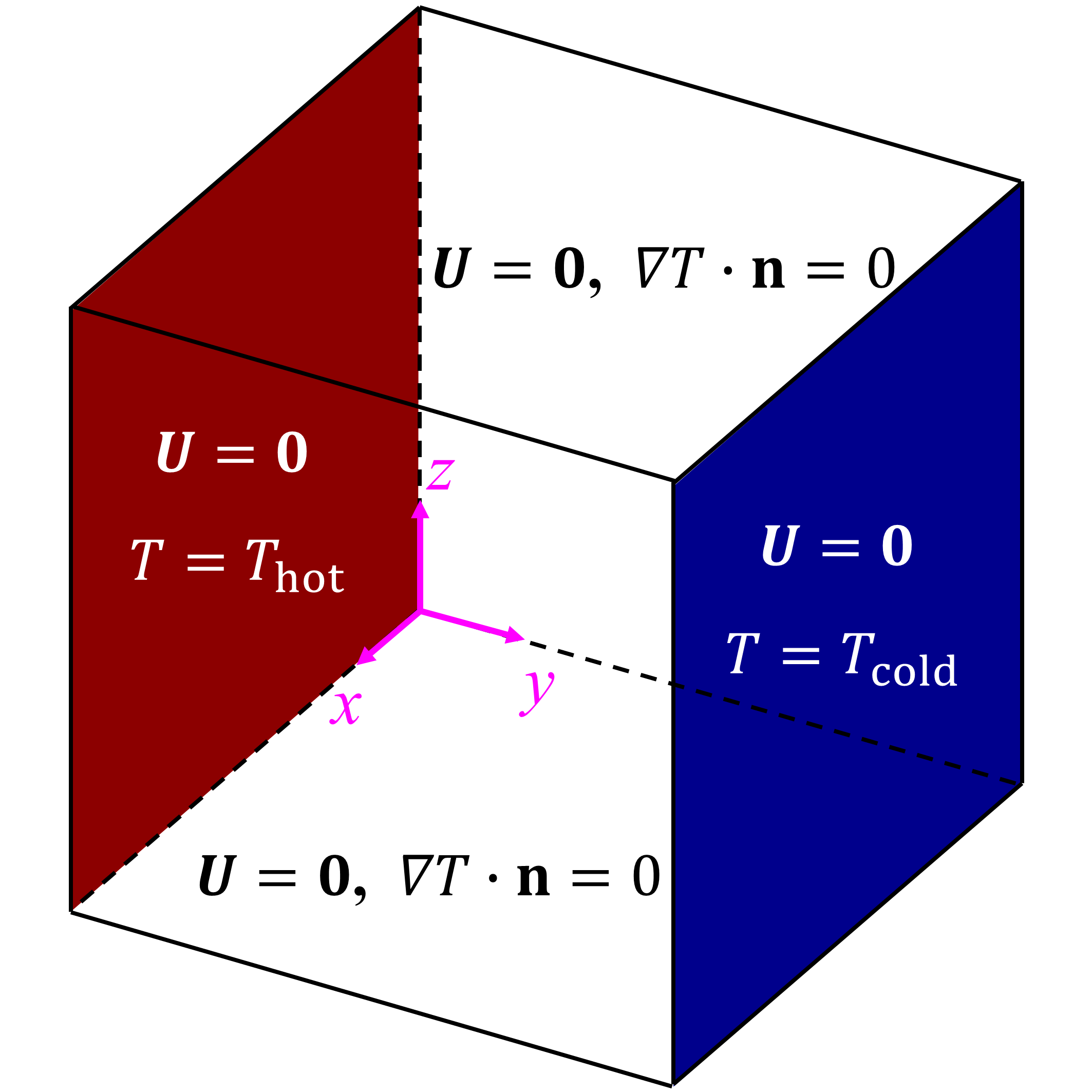}
  \caption{Illustration of the side-heated convection cell.}\label{Fig1}
\end{figure}

Figure 2 shows the temperature fields obtained on grid $N_{x}\times N_{y} \times N_{z} = 257^{3}$ at Ra = $10^{6}$ and $10^{7}$.
The left-hand side is the isothermal surface in the whole cell, while the right-hand side is the temperature cross section along the $x = 0.5$ plane.
At these two high Rayleigh numbers, thin boundary layers exist near isothermal walls; while the temperature stratification is near-linear in the interior region.
In addition, the temperature profiles in the $x = 0.5$ plane generally agrees with prior 2D simulations (see Fig. 10 in our previous work \cite{xu2017accelerated});
while 3D variations of temperature isothermal surface can be observed near the $x = 0$ and $x = 1$ end walls.
\begin{figure}[H]
  \centering
  \includegraphics[width=12cm]{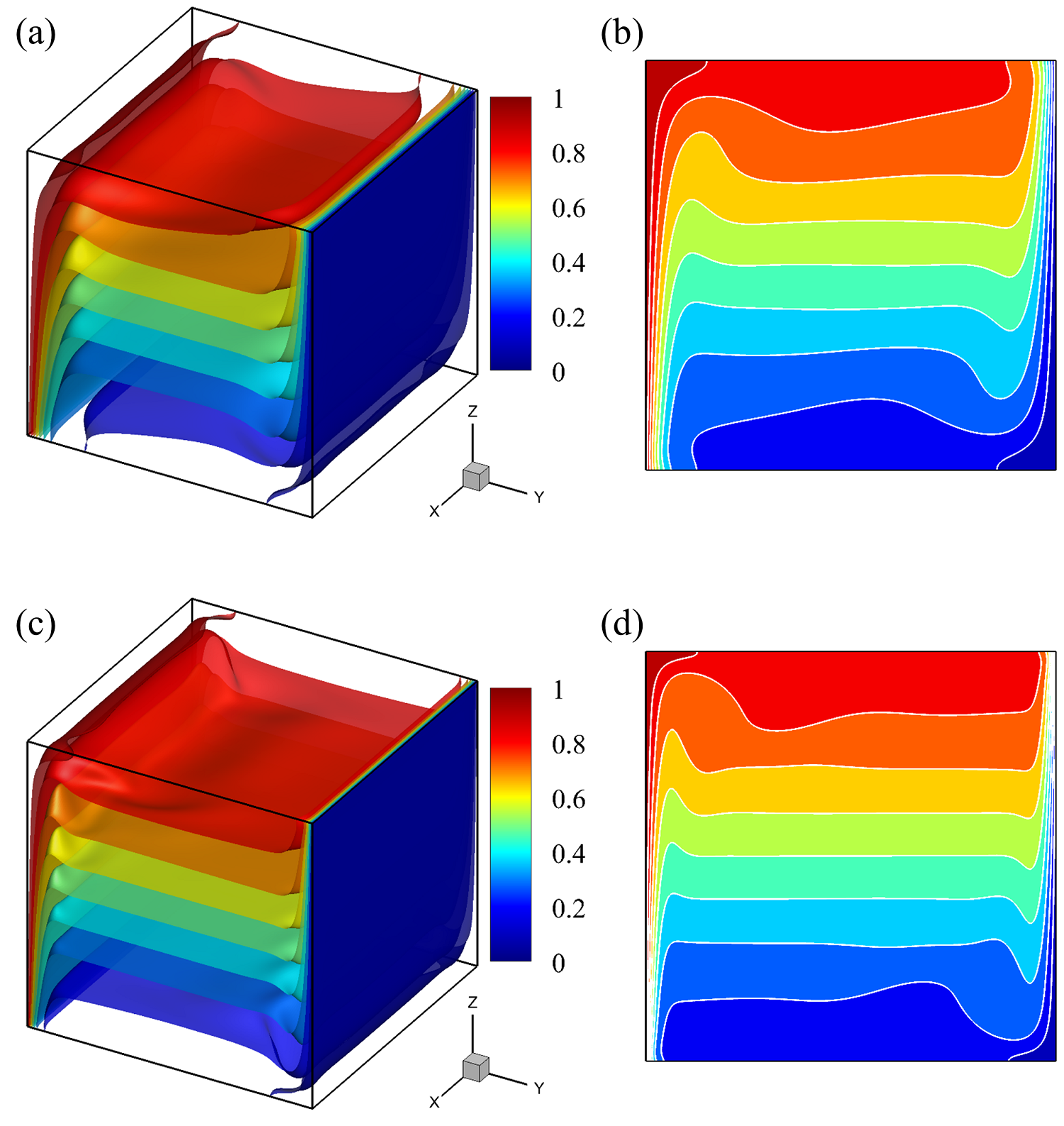}
  \caption{Temperature fields in side-heated convection at (a-b) Ra = $10^{6}$, (c-d) Ra = $10^{7}$; (a, c) isothermal surface in the whole cell, (b, d) cross section at the x = 0.5 plane.}\label{Fig2}
\end{figure}

To provide quantitative results, we first measure some hydrodynamic quantities, including the maximum horizontal velocity $v_{\max}$ at the vertical centerline of the midplane (e.g., $y = 0.5$ line at $x = 0.5$ plane), and its corresponding location $z$;
the maximum vertical velocity $w_{\max}$ at the horizontal centerline of the midplane (e.g., $z = 0.5$ line at $x = 0.5$ plane), and its corresponding location $y$. In addition, we calculate the average kinetic energy $E$ of the system as
\begin{equation}
E=\frac{\frac{1}{2}\int_{\Omega}\left\|\mathbf{u}(\mathbf{x})\right\|^{2}d\mathbf{x}}{\int_{\Omega}d\mathbf{x}}
=\frac{\frac{1}{2}\sum_{i}\left\|\mathbf{u}(\mathbf{x}_{i})\right\|^{2}}{N_{x}N_{y}N_{z}}
\end{equation}
where $\Omega$ is the entire flow domain.
The convergence behaviors of these hydrodynamic quantities are tabulated in Table \ref{Table1}.
We also provide existing data as comparison, such as
Fusegi et al. \cite{fusegi1991numerical} using control-volume based finite different method with strongly implicit scheme to accelerate convergence;
Tric et al. \cite{tric2000first} using pseudo-spectral Chebyshev algorithm based on the projection-diffusion method;
Wang et al. \cite{wang2017numerical} using discrete unified gas-kinetic scheme;
Chen et al. \cite{chen2018high} using high-order simplified thermal lattice Boltzmann method.
It should be noted that in the work of Tric \cite{tric2000first}, the velocity is normalized by $\kappa/L_{0}$, as opposed to $\sqrt{g\beta_{T}L_{0}\Delta_{T}}$  adopted in the present work,
thus values of velocity $\mathbf{u}$ in their work have been divided by $\sqrt{\text{Ra}\cdot\text{Pr}}$  for the convenience of direct comparison.
In addition, in the work of Fusegi et al. \cite{fusegi1991numerical}, Wang et al. \cite{wang2017numerical}, and Chen et al. \cite{chen2018high}, the hot and cold walls are set at $x = 1$ and $x = 0$ planes, respectively;
in the work of Tric et al. \cite{tric2000first}, the hot and cold walls are set at $y = 0.5$ and $y = -0.5$ planes, respectively.
These geometry settings are not identical with present work, where the hot and cold walls are set at $y = 0$ and $y = 1$, respectively.
Thus, the values of the velocity components and its corresponding position have also gone through coordinate transformation.

\begin{small}
\begin{table}[H]
  \centering
  \caption{Convergence behaviors of hydrodynamic quantities.}\label{Table1}
  \begin{tabular}{ccccccccccc}
  \hline
    Ra & Ref. & Mesh size & $v_{\max}$ & $z$ & $w_{\max}$ & $y$ & $E\times 10^{3}$ \\
  \hline
    $10^{6}$ & Fusegi \cite{fusegi1991numerical}  & $62^{3}$                  & 0.08416 & 0.8557 & 0.2588  & 0.0331 & - \\
             & Tric \cite{tric2000first}          & $81^{3}$                  & 0.08096 & 0.8536 & 0.25821 & 0.0331 & - \\
             & Wang \cite{wang2017numerical}      & $50^{3}$                  & 0.0816  & 0.8597 & 0.2556  & 0.0347 & - \\
             & Chen \cite{chen2018high}           & $101\times 51 \times 101$ & 0.080   & 0.860  & 0.257   & 0.040  & - \\
             & Present & $81^{3}$                  & 0.08056 & 0.8580       & 0.25437       & 0.0432 & 3.3346 \\
             & Present & $129^{3}$                 & 0.08091 & 0.8566       & 0.25753       & 0.0349 & 3.3280 \\
             & Present & $161^{3}$                 & 0.08099 & 0.8540       & 0.25729       & 0.0404 & 3.3265 \\
             & Present & $257^{3}$                 & 0.08107 & 0.8541       & 0.25836       & 0.0370 & 3.3248 \\
  \hline
    $10^{7}$ & Tric \cite{tric2000first}           & $111^{3}$                 & 0.05813 & 0.8716 & 0.25994 & 0.0194 & - \\
             & Wang \cite{wang2017numerical}       & $200^{3}$                 & 0.0558  & 0.8831 & 0.2590  & 0.0233 & - \\
             & Chen \cite{chen2018high}            & $121\times 51 \times 121$ & 0.0585  & 0.8750 & 0.2606  & 0.0199 & - \\
             & Present & $81^{3}$                  & 0.05410 & 0.8951 & 0.25865 & 0.0185 & 1.8418 \\
             & Present & $129^{3}$                 & 0.05671 & 0.8798 & 0.26054 & 0.0194 & 1.8365 \\
             & Present & $161^{3}$                 & 0.05730 & 0.8789 & 0.26184 & 0.0217 & 1.8345 \\
             & Present & $257^{3}$                 & 0.05789 & 0.8774 & 0.26181 & 0.0214 & 1.8322 \\
  \hline
  \end{tabular}
\end{table}
\end{small}

We then measure Nusselt numbers to quantify the heat transfer process.
We consider the mean Nusselt number $\text{Nu}_{\text{mean}}$ at the $x = 0.5$ midplane along the hot wall ($y = 0$) and the cold wall ($y = 1$);
the overall Nusselt number $\text{Nu}_{\text{overall}}$ along the hot and cold walls. Here, $\text{Nu}_{\text{mean}}(x)$ and $\text{Nu}_{\text{overall}}$ are defined as
\begin{equation}
\text{Nu}_{\text{mean}}(x)=-\int_{0}^{1} \frac{\partial T(x,z)}{\partial y} \bigg|_{y=0 \ \text{or} \ y=1} dz
\end{equation}
\begin{equation}
\text{Nu}_{\text{overall}}=-\int_{0}^{1} \int_{0}^{1} \frac{\partial T(x,z)}{\partial y} \bigg|_{y=0 \ \text{or} \ y=1} dx \ dz
\end{equation}
The convergence behavior of these Nusselt numbers are tabulated in Table \ref{Table2}.
In addition, the asymptotic values $f_{\infty}$ are used as the reference values to compute the relative error, which are then used to estimate the order of accuracy $n$ for LB simulation.
At $\text{Ra}=10^{7}$, results obtained at coarse mesh size of $81^{3}$ do not fit well with the interpolating polynomial, and they have been excluded from computing asymptotic values.
Overall, the present thermal LB model has an approximate second-order spatial accuracy.

\begin{table}[H]
  \centering
  \caption{Convergence behaviors of Nusselt numbers.}\label{Table2}
  \begin{tabular}{ccccccccccc}
  \hline
  Ra & Ref. & Mesh size & $\text{Nu}_{\text{mean}}$  & $\text{Nu}_{\text{mean}}$ & $\text{Nu}_{\text{overall}}$ & $\text{Nu}_{\text{overall}}$ \\
     &      &           & $y=0$                      & $y=1$                     & $y=0$                        & $y=1$ \\
  \hline
  $10^{6}$  & Fusegi \cite{fusegi1991numerical} & $62^{3}$                      & 9.012   & -       & 8.770   & - \\
            & Tric \cite{tric2000first}         & $81^{3}$                      & 8.8771  & -       & 8.6407  & - \\
            & Wang \cite{wang2017numerical}     & $50^{3}$                      & 8.7795  & -       & 8.5428  & - \\
            & Chen \cite{chen2018high}          & $101 \times 51 \times 101$    & 9.072   & -       & 8.741   & - \\
            & Present                           & $81^{3}$                      & 8.99850  & 8.99333  & 8.75450  & 8.75405 \\
            & Present                           & $129^{3}$                     & 8.91688  & 8.91467  & 8.67775  & 8.67746 \\
            & Present                           & $161^{3}$                     & 8.89886  & 8.89744  & 8.66075  & 8.66060 \\
            & Present                           & $257^{3}$                     & 8.88054  & 8.87994  & 8.64345  & 8.64342 \\

            &                                   & $f_{\infty}$                      & 8.8735  & 8.8728  & 8.6369  & 8.6364  \\
            &                                   & $n$                           & 2.49    & 2.45    & 2.49    & 2.44    \\
  \hline
  $10^{7}$  & Tric \cite{tric2000first}         & $111^{3}$                     & 16.5477 & -       & 16.3427 & - \\
            & Wang \cite{wang2017numerical}     & $200^{3}$                     & 16.4153 & 16.3909 & 16.2112 & 16.1872 \\
            & Chen \cite{chen2018high}          & $121 \times 51 \times 121$    & 16.457  & -       & 16.604  & - \\
            & Present                           & $81^{3}$                      & 17.26877$^{*}$ & 17.31377$^{*}$ & 17.03864$^{*}$ & 17.09483$^{*}$ \\
            & Present                           & $129^{3}$                     & 16.85522       & 16.85465       & 16.64342       & 16.64642 \\
            & Present                           & $161^{3}$                     & 16.73787       & 16.73588       & 16.52963       & 16.52950 \\
            & Present                           & $257^{3}$                     & 16.60871       & 16.60782       & 16.40322       & 16.40285 \\

            &                                   & $f_{\infty}$                      & 16.5204        & 16.5301        & 16.3124        & 16.3237  \\
            &                                   & $n$                           & 1.93           & 2.07           & 1.88           & 2.04   \\
  \hline
  \end{tabular}
\end{table}

In Tables \ref{Table1} and \ref{Table2}, the results given by Wang et al. \cite{wang2017numerical} at Ra = $10^{7}$ were time-averaged quantities, indicating their simulations did not converge to steady states;
while Tric et al. \cite{tric2000first} and Chen et al. \cite{chen2018high} mentioned natural convection in such a configuration enters unsteady flow regime at Rayleigh number beyond $10^{7}$.
Here, we present convergence histories of velocity $\mathbf{u}$ in Fig. 3.
With the present LB model and the four mesh sizes of $81^{3}$, $129^{3}$, $161^{3}$ and $257^{3}$, our simulations were able to reach residual errors down to $10^{-9}$;
similar convergence histories of temperature $T$ were also observed, but not shown here for clarity.
It is worth mentioning in numerical investigations, the bifurcation critical number depends on the formulation, numerical method, and choice of grid.
Even for the canonical lid-driven cavity problem that only considers incompressible isothermal flows, different researchers presented various first bifurcation critical Reynolds numbers \cite{xu2017accelerated,suman2019grid}.
\begin{figure}[H]
  \centering
  \includegraphics[width=12cm]{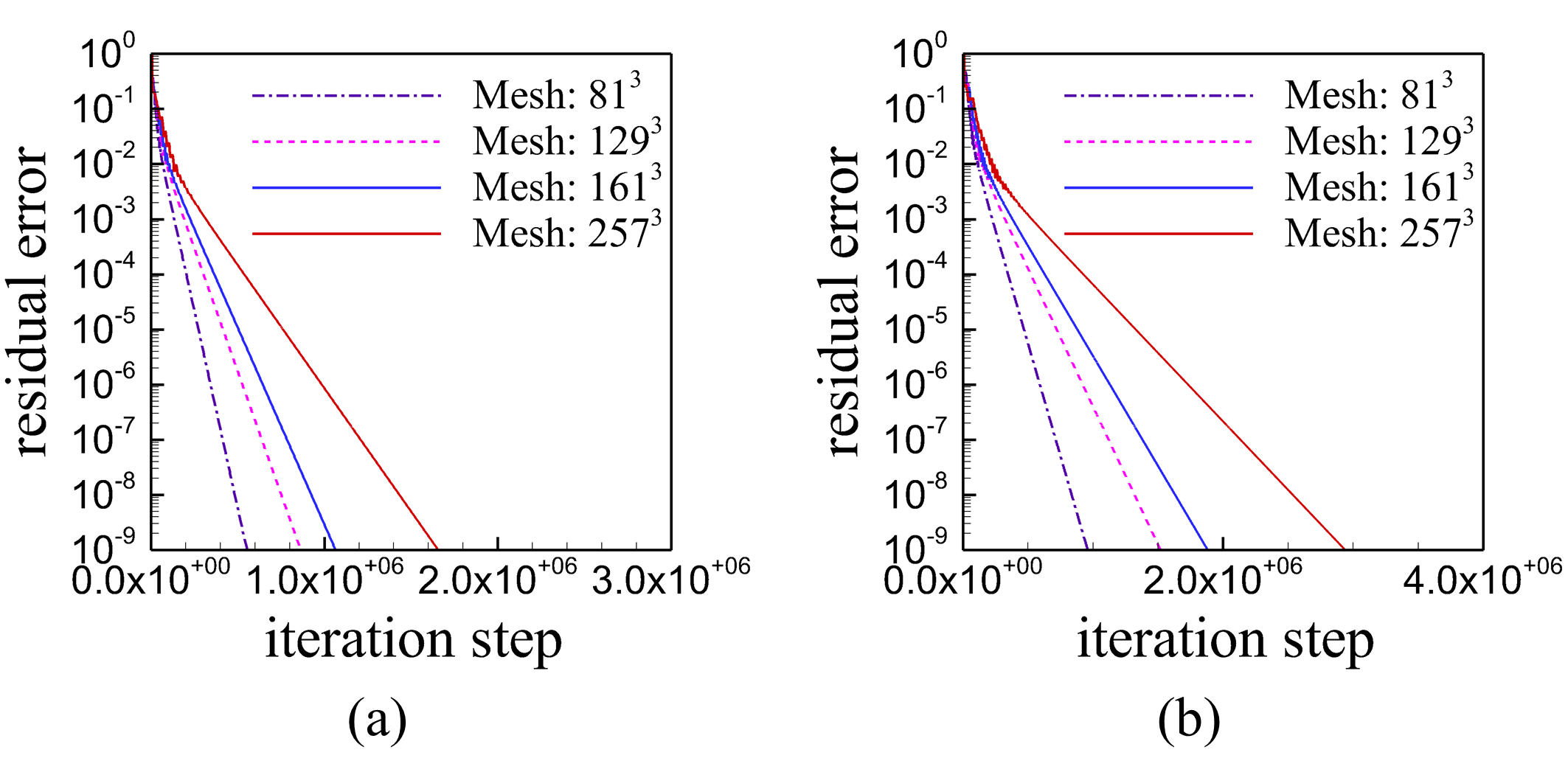}
  \caption{Convergence history of velocity $\mathbf{u}$ at (a) Ra = $10^{6}$ and (b) Ra = $10^{7}$.}\label{Fig3}
\end{figure}

We further show the $y$ variation of the Nusselt number averaged over $x$-$z$ plane in Fig. 4.
Here, the $x$-$z$ plane averaged Nusselt number is defined as
\begin{equation}
\text{Nu}(y)=\int_{0}^{1}\int_{0}^{1}\left(vT\sqrt{\text{Ra}\text{Pr}}-\frac{\partial T}{\partial y} \right)dx \ dz
\end{equation}
We can see from Fig. 4, the Nusselt number oscillates near the hot or cold walls ($y = 0$ or $y = 1$), which is due to lack of mesh resolution.
When increasing mesh sizes, the amplitude of this small variation will decrease, and the Nusselt number will converge to a constant.
\begin{figure}[H]
  \centering
  \includegraphics[width=12cm]{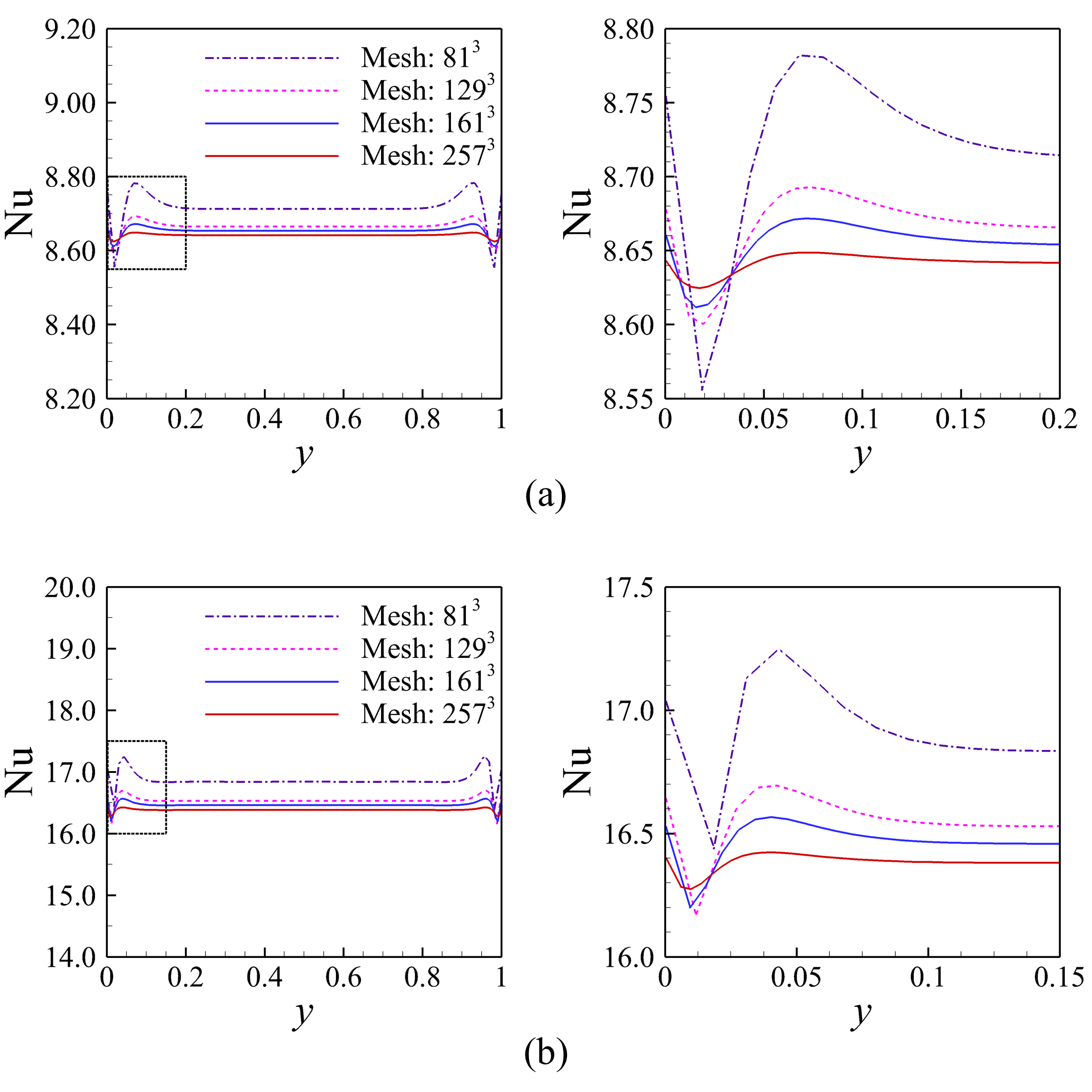}
  \caption{The Nusselt number averaged over $x$-$z$ plane as a function of $y$ at (a) Ra = $10^{6}$ and (b) Ra = $10^{7}$; subfigures in the right column are zoomed-in views of corresponding subfigures in the left column.}\label{Fig4}
\end{figure}

Since the lattice Boltzmann method intrinsically solves weakly compressible Navier-Stokes equations, to directly quantify the compressibility effect, we then compute the root-mean-square (rms) density fluctuation and the rms velocity divergence as
\begin{equation}
\sqrt{\langle (\delta \rho)^{2} \rangle} = \left[ \frac{\sum_{i}(\rho_{i}-\rho_{0})^{2}}{N_{x}N_{y}N_{z}} \right]^{1/2}
\end{equation}
\begin{equation}
\sqrt{\langle (\nabla \cdot \mathbf{u})^{2} \rangle}=\left[ \frac{\sum_{i}(\nabla \cdot \mathbf{u}_{i})^{2}}{N_{x}N_{y}N_{z}} \right]^{1/2}
\end{equation}
We can see from Table 3, both the Rayleigh number and mesh sizes have little effects on the rms density fluctuation; while the rms velocity divergence decreases when increasing the mesh sizes or decreasing the Rayleigh number.
The dependence of rms velocity divergence on mesh sizes or Rayleigh number can be explained as follows.
In the present LB model, the relaxation parameters were adjusted following the principles in TRT model, i.e., $s_{e}=s_{\varepsilon}=s_{\nu}=s_{\pi}=1/\tau_{f}$,  $s_{q}=s_{m}=8(2\tau_{f}-1)/(8\tau_{f}-1)$, which leads to the bulk viscosity equal to shear viscosity as  $\zeta=c_{s}^{2}(s_{e}^{-1}-0.5)\delta_{t}=c_{s}^{2}(s_{\nu}^{-1}-0.5)\delta_{t}=\nu$.
At fixed Rayleigh number, increasing the mesh sizes leads to larger shear viscosity and bulk viscosity, thus resulting in stronger dissipation of modes related to compressibility, and smaller rms velocity divergence; at fixed mesh size, increasing the Rayleigh number leads to smaller bulk viscosity, resulting in larger rms velocity divergence.
On the other hand, we notice that the simulation results reported by Ostilla-Monico et al. \cite{ostilla2015multiple}, who used the finite difference method coupling with multiple-resolution strategy to directly solve the Navier-Stokes equations, also show non-solenoidal velocity field with small residual divergence of $O(10^{-3})$.
So far, this small magnitude of residual divergence has not resulted in apparent problems when simulating incompressible thermal convective flows, even for flows in turbulent flow regime.

\begin{table}[H]
  \centering
  \caption{The root-mean-square density fluctuation and velocity divergence.}\label{Table3}
  \begin{tabular}{ccccccccccc}
  \hline
  Ra        & Mesh size & rms density fluctuation & rms velocity divergence \\
  \hline
  $10^{6}$  & $81^{3}$  & $1.4793 \times 10^{-3}$ & $1.994 \times 10^{-2}$ \\
            & $129^{3}$ & $1.4794 \times 10^{-3}$ & $8.779 \times 10^{-3}$ \\
            & $161^{3}$ & $1.4794 \times 10^{-3}$ & $5.722 \times 10^{-3}$ \\
            & $257^{3}$ & $1.4795 \times 10^{-3}$ & $2.228 \times 10^{-3}$ \\
  \hline
  $10^{7}$  & $81^{3}$  & $1.4770 \times 10^{-3}$ & $4.183 \times 10^{-2}$ \\
            & $129^{3}$ & $1.4788 \times 10^{-3}$ & $2.295 \times 10^{-2}$ \\
            & $161^{3}$ & $1.4791 \times 10^{-3}$ & $1.617 \times 10^{-2}$ \\
            & $257^{3}$ & $1.4795 \times 10^{-3}$ & $7.132 \times 10^{-3}$ \\
  \hline
  \end{tabular}
\end{table}

\section{Turbulent thermal convection in Rayleigh-B\'enard convection cell}

\begin{figure}[H]
  \centering
  \includegraphics[width=6cm]{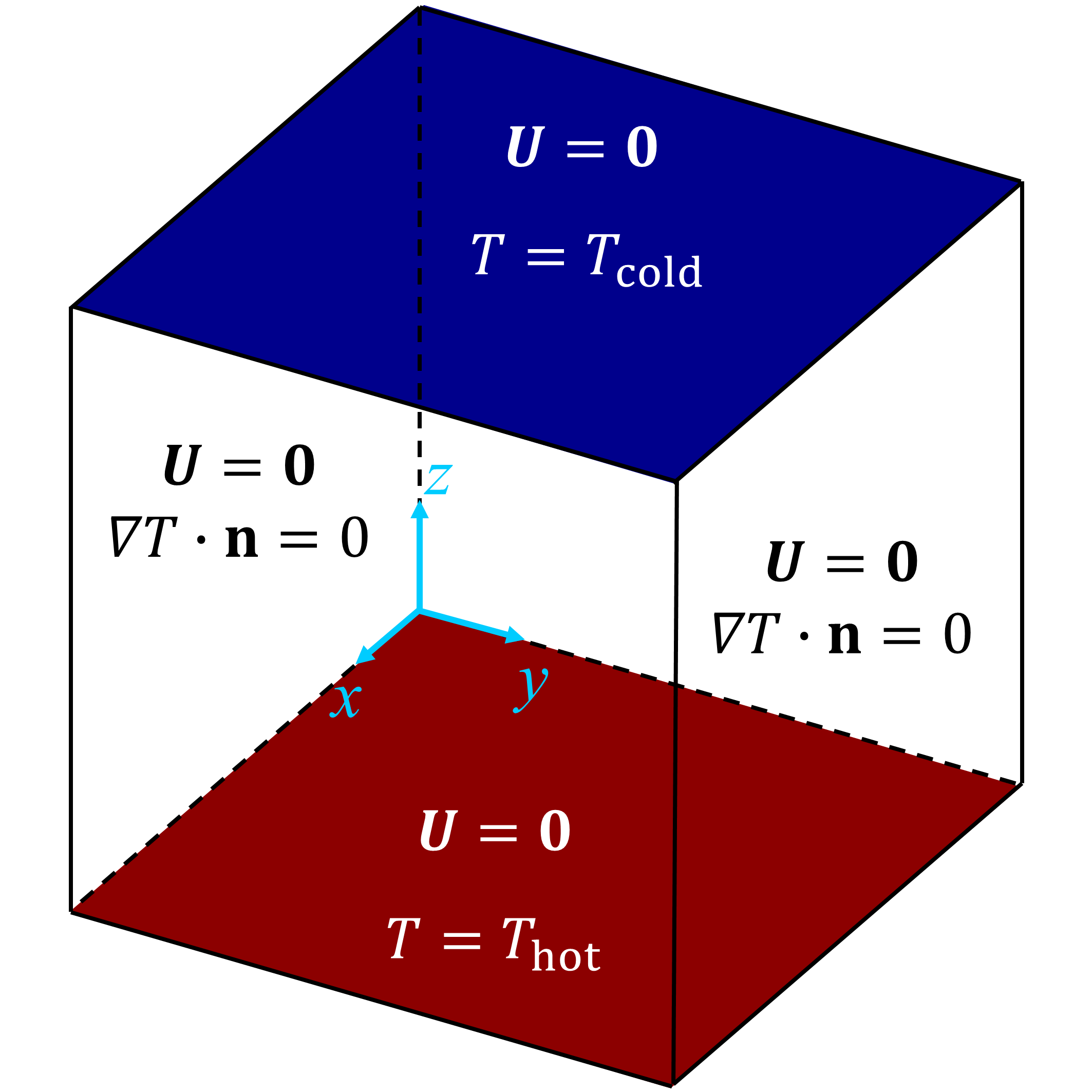}
  \caption{Illustration of the Rayleigh-B\'enard convection cell.}\label{Fig5}
\end{figure}

The flow configuration for the RB cell is shown in Fig. \ref{Fig5}.
The top and bottom walls are kept at constant cold and hot temperature, respectively;
while the other four vertical walls are adiabatic.
All six walls impose no-slip velocity boundary condition.
The dimension of the cell is $L \times D \times H$, and we set $L=D=H$ in this work.
Simulation results are provided at Rayleigh numbers of $10^{6}$, $3\times 10^{6}$, and $10^{7}$, and Prandtl numbers of 0.7 and 7.
The Mach number is fixed as 0.1.
The simulation protocol is as follows: first check whether statistically stationary state has reached in every 100 dimensionless time units;
after that check whether statistically converge state has reached in every 200 dimensionless time units.

To measure global strength of the convection, the Reynolds number based on root-mean-square (rms) velocity is defined as
\begin{equation}
\text{Re}^{\text{rms}}=\frac{\sqrt{\langle (u^{2}+v^{2}+w^{2})\rangle_{V,t}}H}{\nu}
\end{equation}
where $\langle \cdot \rangle_{V,t}$ denotes an ensemble average over the whole cell and over time.
Similarly, the Reynolds number based on rms vertical velocity (i.e., parallel to gravity) is defined as
\begin{equation}
\text{Re}_{w}^{\text{rms}}=\frac{\sqrt{\langle w^{2}\rangle_{V,t}}H}{\nu}
\end{equation}

To measure global heat transport, the volume average Nusselt number ($\text{Nu}_{\text{vol}}$) is calculated as \cite{kerr1996rayleigh,verzicco2003numerical}
\begin{equation}
\text{Nu}_{\text{vol}}=1+\sqrt{\text{Pr}\text{Ra}}\langle wT \rangle_{V,t}
\end{equation}
Meanwhile, since no-slip velocity is imposed on the top and bottom walls, the average Nusselt number over top and bottom walls can be calculated as
\begin{equation}
\text{Nu}_{\text{wall}}=-\frac{1}{2}\left( \langle \partial_{z}T \rangle_{\text{top},t}+\langle \partial_{z}T \rangle_{\text{bottom},t} \right)
\end{equation}
where $\langle \cdot \rangle_{\text{top}}$  and $\langle \cdot \rangle_{\text{bottom}}$ denotes an ensemble average over the top and bottom walls, respectively.
In addition, by averaging the equations of motion, we can define another two Nusselt numbers related with global averages of kinetic and thermal energy dissipation rates as \cite{shraiman1990heat,siggia1994high}
\begin{equation}
\text{Nu}_{\text{kin}}=1+\sqrt{\text{Ra}\text{Pr}}\langle \varepsilon_{u} \rangle_{V,t}
\end{equation}
\begin{equation}
\text{Nu}_{\text{th}}=\sqrt{\text{Ra}\text{Pr}}\langle \varepsilon_{T} \rangle_{V,t}
\end{equation}
where the kinetic and thermal energy dissipation rates are given by
\begin{equation}
\varepsilon_{u}(\mathbf{x},t)=\frac{1}{2}\nu \sum_{ij}\left[ \frac{\partial u_{j}(\mathbf{x},t)}{\partial x_{i}}+\frac{\partial u_{i}(\mathbf{x},t)}{\partial x_{j}} \right]^{2}
\end{equation}
\begin{equation}
\varepsilon_{T}(\mathbf{x},t)=\kappa \sum_{i}\left[ \frac{\partial T(\mathbf{x},t)}{\partial x_{i}} \right]^{2}
\end{equation}
The above rigorous relations further form the backbone of the Grossmann-Lohse (GL) theory of turbulent heat transfer \cite{grossmann2000scaling,grossmann2004fluctuations}.

Table \ref{Table4} tabulates the values for Reynolds and Nusselt numbers obtained from the present simulations.
If the direct numerical simulation of RB convection is well resolved and statistically convergent, the above definitions of Nusselt numbers should give results agree with each other. Here, the volume averaged Nusselt number $\text{Nu}_{\text{vol}}$ is chosen as the reference value to calculate its relative differences with other Nusselt numbers, and the results (denoted by 'diff.') are included in the bracket in corresponding columns.
From Table \ref{Table4}, we can see the differences are within 1\%, indicating that Nusselt numbers show good consistency with each other.
We further fit the data to obtain scaling relations of Reynolds number and Nusselt number versus Rayleigh number using power-law relations.
For the $\text{Ra} \sim \text{Re}$ scaling, we have
$\text{Re}^{\text{rms}}=0.209\text{Ra}^{0.499\pm 0.004}$  and  $\text{Re}_{w}^{\text{rms}}=0.154\text{Ra}^{0.496\pm 0.001}$ at Pr = 0.7,
while $\text{Re}^{\text{rms}}=0.016\text{Ra}^{0.532\pm 0.017}$ and $\text{Re}_{w}^{\text{rms}}=0.013\text{Ra}^{0.529\pm 0.025}$  at Pr = 7,
which agree well with previous studies that $\text{Re}$ proportional to $\text{Ra}^{1/2}$ \cite{van2013comparison}.
For the $\text{Ra} \sim \text{Nu}$ scaling, we have
$\text{Nu}=0.153\text{Ra}^{0.289\pm 0.0003}$ at Pr = 0.7,
while $\text{Nu}=0.158\text{Ra}^{0.287\pm 0.018}$ at Pr = 7,
which agree well with previous studies that $\text{Nu}$ proportional to $\text{Ra}^{2/7}$ \cite{castaing1989scaling,wagner2012boundary}.

\begin{table}[H]
  \centering
  \caption{The Reynolds and Nusselt numbers in Rayleigh-B\'enard convection.}\label{Table4}
\begin{small}
  \begin{tabular}{cccccccccccc}
  \hline
  Ra & Pr & $\text{Re}^{\text{rms}}$ & $\text{Re}_{w}^{\text{rms}}$ & $\text{Nu}_{\text{vol}}$ & $\text{Nu}_{\text{wall}}$ (diff.) & $\text{Nu}_{\text{kin}}$ (diff.)  & $\text{Nu}_{\text{th}}$ (diff.)  \\
  \hline
  $10^{6}$          & 0.7 & 208.80 & 145.58 & 8.33   & 8.35  (0.22\%)  & 8.24  (1.06\%)  & 8.26  (0.88\%) \\
  $3 \times 10^{6}$ & 0.7 & 357.11 & 249.96 & 11.46  & 11.48 (0.23\%)  & 11.35 (0.89\%)  & 11.37 (0.72\%) \\
  $10^{7}$          & 0.7 & 654.86 & 454.92 & 16.22  & 16.27 (0.30\%)  & 16.07 (0.96\%)  & 16.10 (0.78\%) \\
  \hline
  $10^{6}$          & 7   & 26.13  & 20.26  & 8.49   & 8.52  (0.33\%)  & 8.46  (0.42\%)  & 8.44  (0.57\%) \\
  $3 \times 10^{6}$ & 7   & 44.56  & 33.65  & 11.12  & 11.14 (0.24\%)  & 11.08 (0.35\%)  & 11.06 (0.52\%) \\
  $10^{7}$          & 7   & 86.41  & 65.71  & 16.16  & 16.19 (0.16\%)  & 16.11 (0.35\%)  & 16.03 (0.80\%) \\
  \hline
  \end{tabular}
\end{small}
\end{table}

Sufficiently resolved simulations would give converging Nusselt numbers, but not vice versa.
For example, Kooij et al. \cite{kooij2018comparison} observed ripples in instantaneous snapshots of temperature fields near sharp gradients when the simulation is under-resolved, even though the Nusselt numbers from the simulations look reasonable.
Thus, we also check whether the grid spacing $\Delta_{g}$ and time interval $\Delta_{t}$ is properly resolved by comparing with the Kolmogorov and Batchelor scales.
Here, the Kolmogorov length scale is estimated by the global criterion $\eta=H\text{Pr}^{1/2}/[\text{Ra}(\text{Nu}-1)]^{1/4}$, the Batchelor length scale is estimated by $\eta_{B}=\eta \text{Pr}^{-1/2}$, and the Kolmogorov time scale is estimated as $\tau_{\eta}=\sqrt{\nu/\langle \varepsilon_{u} \rangle}=\sqrt{\text{Pr}/(\text{Nu}-1)}$.
From Table 5, we can see that grid spacings satisfy $\max \left( \Delta_{g}/\eta, \Delta_{g}/\eta_{B} \right) \le 0.52$, which ensures the spatial resolution.
In addition, the time intervals are $\Delta_{t} \le 0.00145 \tau_{\eta}$, thus guaranteeing adequate temporal resolution.
However, such a fine temporal resolution is the result of intrinsic defects in LB time marching scheme, the small time steps was not adopted on purpose.
Specifically, the Courant-Friedrichs-Lewy (CFL) number in LB method can be calculated as $\text{CFL}_{\text{LB}}=dt/dx=\big(\delta_{t}/\sqrt{L_{0}/(g\beta_{T}\Delta_{T})}\big)/\big(\delta_{x}/L_{0}\big)=\sqrt{L_{0}g\beta_{T}\Delta_{T}}=\text{Ma}\cdot c_{s}\approx 0.0577$, where $\delta_{x}=\delta_{t}=1$, $c_{s}=1/\sqrt{3}$, and Ma = 0.1 have been used in our simulations.
In conventional numerical methods that directly solve the Navier-Stokes equations, the CFL numbers can be five to six times larger, leading to larger time interval.
On the other hand, it should be noted that the LB method does not require to solve the time consuming pressure Poisson equation, which saves the computational cost compared with the conventional Navier-Stokes solvers.
Thus, a compressive compression of the overall computing efficiency between different numerical methods is needed in the future.
In Table 5, we also estimate the number of grid points within the thermal boundary layer, where $N^{th}_{\text{BL}}\approx H/(2\text{Nu})$ \cite{shishkina2010boundary}.
Around 8 nodes are used within the thermal boundary layers in all the cases.
To make sure statistically stationary state has been reached and the initial transient effects are washed out, we first simulate a time period of at least $500t_{f}$.
After that, an additional averaging time $t_{\text{avg}}$ of at least $200t_{f}$ (one case even with $1800t_{f}$) are simulated to reach the statistical convergence state.
Here, $t_{f}$ denotes free-fall time units $t_{f}=\sqrt{H/(g\beta_{T}\Delta_{T})}$.

\begin{table}[H]
  \centering
  \caption{Spatial and temporal resolutions of the simulations.}\label{Table5}
  \begin{tabular}{cccccccccccc}
  \hline
  Ra & Pr & Mesh size & $\Delta_{g}/\eta$ & $\Delta_{g}/\eta_{B}$ & $\Delta_{t}/\tau_{\eta}$ & $N^{th}_{\text{BL}}$ & $t_{\text{avg}}/t_{f}$ \\
  \hline
  $10^{6}$          & 0.7 & $129^{3}$ & 0.48 & 0.40 & $1.45\times 10^{-3}$ & 8  & 1800 \\
  $3 \times 10^{6}$ & 0.7 & $193^{3}$ & 0.46 & 0.39 & $1.16\times 10^{-3}$ & 8  & 400 \\
  $10^{7}$          & 0.7 & $257^{3}$ & 0.52 & 0.43 & $1.05\times 10^{-3}$ & 8  & 400 \\
  \hline
  $10^{6}$          & 7   & $129^{3}$ & 0.15 & 0.41 & $4.63\times 10^{-4}$ & 8  & 200 \\
  $3\times 10^{6}$  & 7   & $193^{3}$ & 0.15 & 0.38 & $3.60\times 10^{-4}$ & 9  & 800 \\
  $10^{7}$          & 7   & $257^{3}$ & 0.16 & 0.43 & $3.31\times 10^{-4}$ & 8  & 800 \\
  \hline
  \end{tabular}
\end{table}

In addition to statistically averaged Reynolds and Nusselt numbers, we show instantaneous flow and temperature structures in Fig. 6.
We can observe hot and cold plumes in mushroom-like shape detaching from both the top and bottom thermal boundary layers of the cell.
In addition, the maximum absolute value of vertical velocity is higher at Pr = 0.7 (Fig. 6c) compared with that at Pr = 7 (Fig. 6d), indicating stronger motion of upward and downward moving fluids at lower Prandtl number.
Fig. 7 further presents logarithmic kinetic energy dissipation fields and logarithmic thermal energy dissipation fields.
Since rising and falling thermal plumes are associated with large amplitudes of both kinetic and thermal energy dissipation rates, intense dissipations occur almost in regions with higher or lower temperature.
\begin{figure}[H]
  \centering
  \includegraphics[width=12cm]{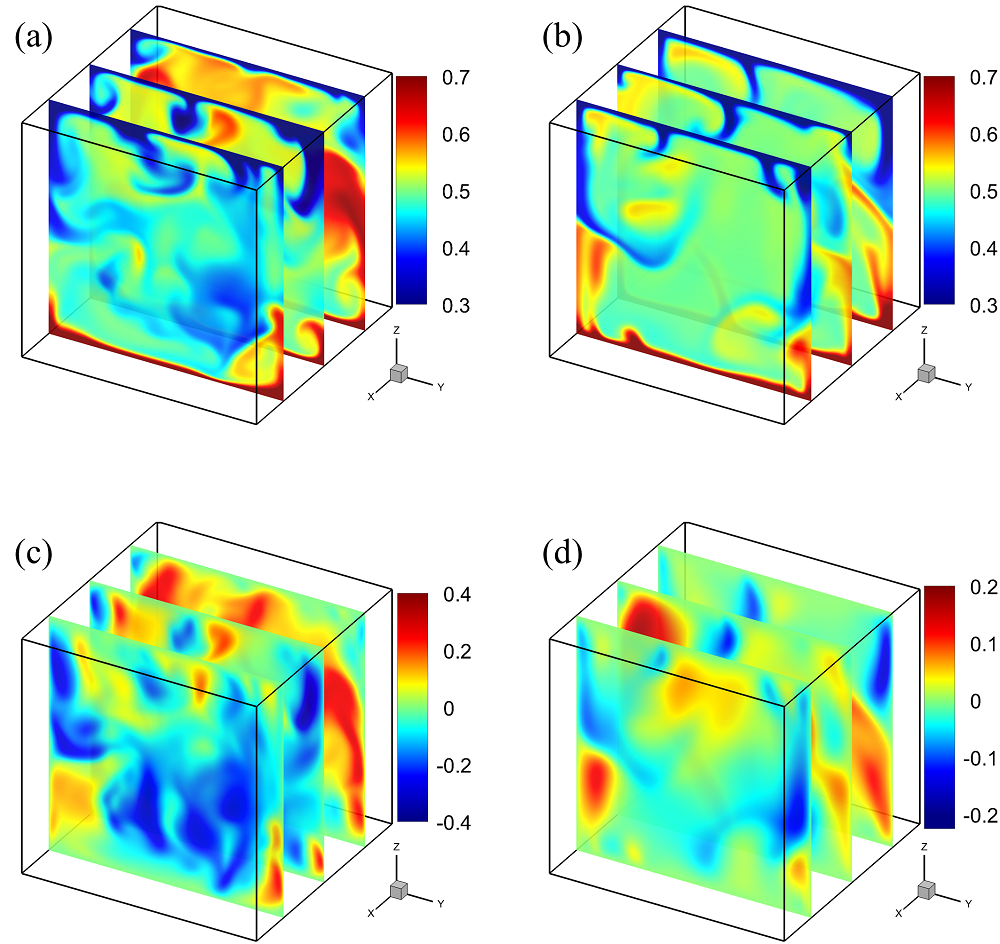}
  \caption{Typical snapshots of the instantaneous (a-b) temperature fields, (c-d) vertical velocity fields at Ra = $10^{7}$, (a, c) Pr = 0.7, and (b, d) Pr = 7.}\label{Fig6}
\end{figure}
\begin{figure}[H]
  \centering
  \includegraphics[width=12cm]{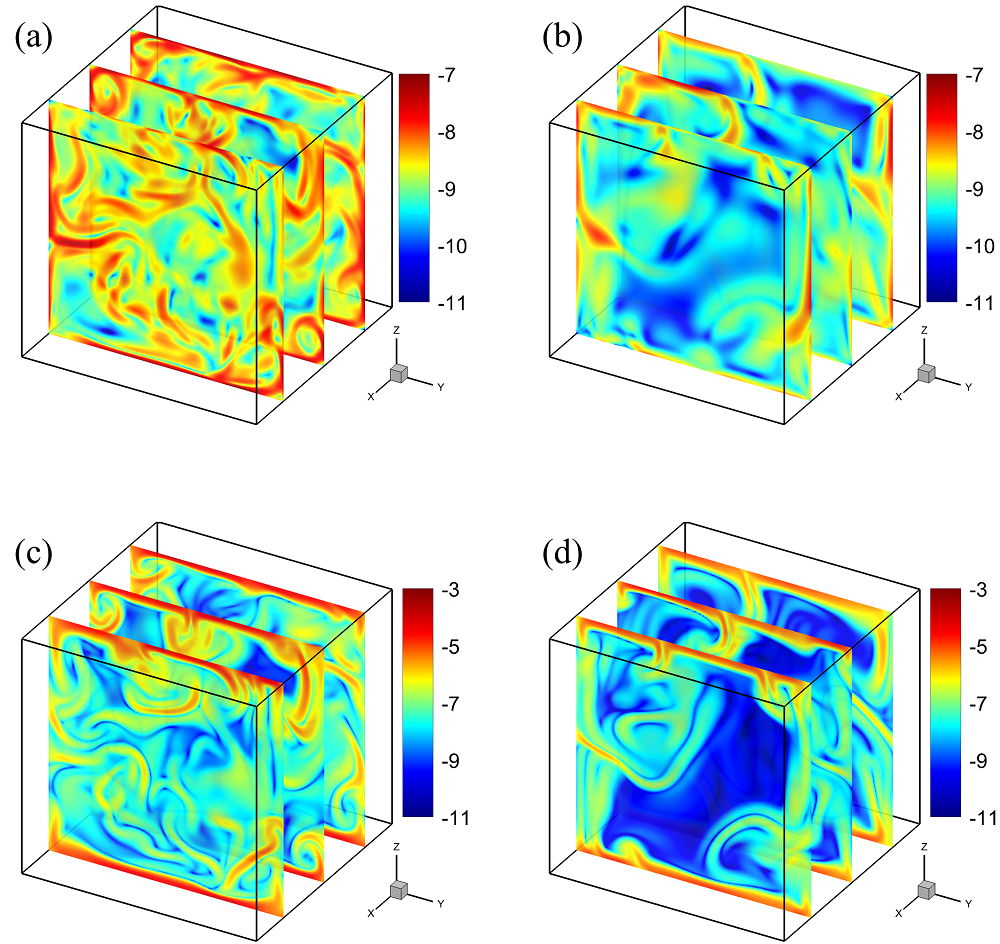}
  \caption{Typical snapshots of the instantaneous (a-b) logarithmic kinetic energy dissipation fields, (c-d) logarithmic thermal energy dissipation fields at Ra = $10^{7}$, (a, c) Pr = 0.7, and (b, d) Pr = 7.}\label{Fig7}
\end{figure}

\begin{figure}[H]
  \centering
  \includegraphics[width=12cm]{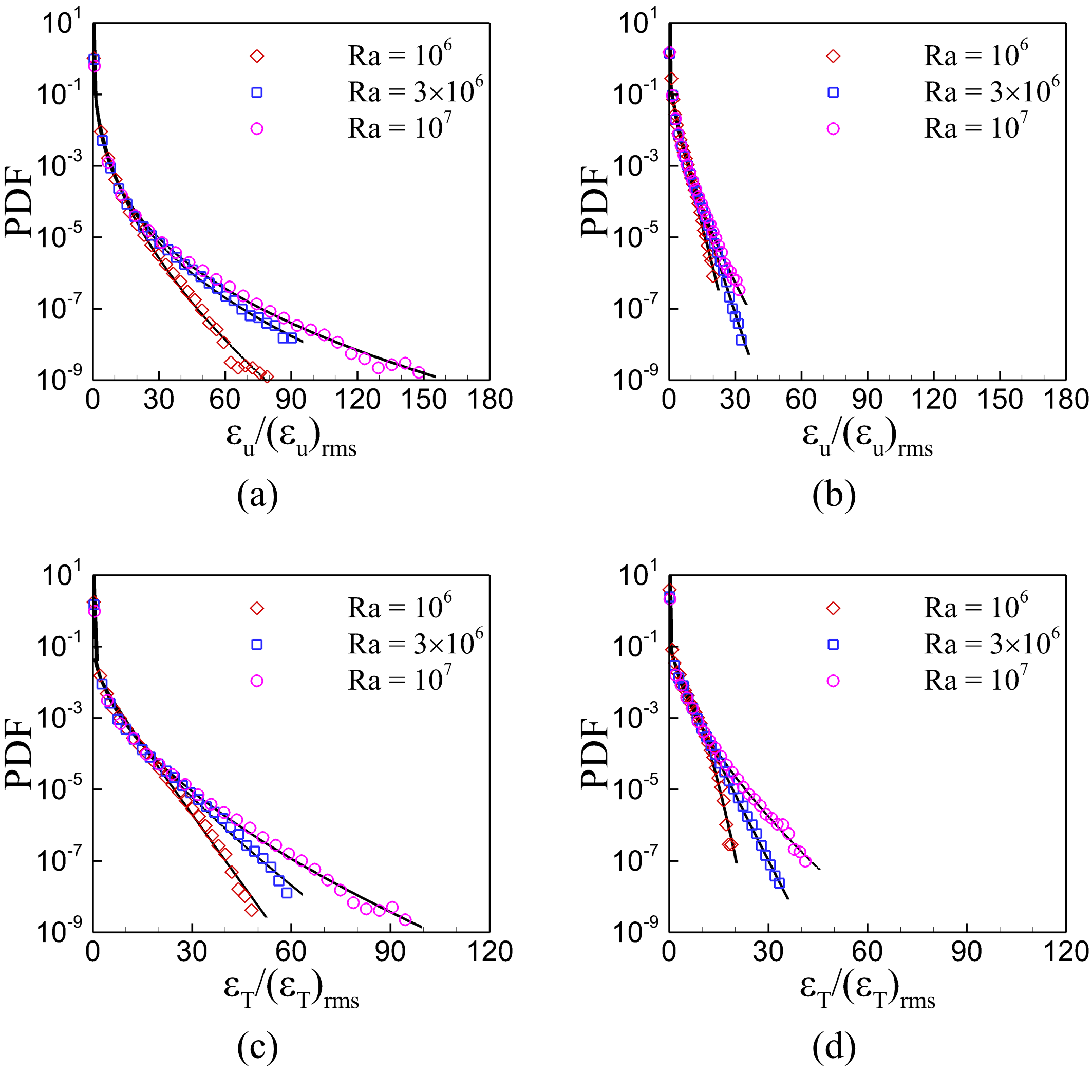}
  \caption{PDFs of kinetic energy dissipation rates at (a) Pr = 0.7, (b) Pr = 7; and thermal energy dissipation rates at (c) Pr = 0.7, (d) Pr = 7.}\label{Fig8}
\end{figure}

The probability density functions (PDFs) of kinetic and thermal energy dissipation rates obtained over the whole cell are shown in Fig. 8.
All data have been normalized with respect to their root-mean-square values, where $(\varepsilon_{u})_{\text{rms}}=\sqrt{\langle \varepsilon_{u}^{2} \rangle_{V,t}}$  and  $(\varepsilon_{T})_{\text{rms}}=\sqrt{\langle \varepsilon_{T}^{2} \rangle_{V,t}}$.
At the same Rayleigh number, decreasing the Prandtl number (e.g., Fig. \ref{Fig8}b versus Fig. \ref{Fig8}a, and Fig. \ref{Fig8}d versus Fig. \ref{Fig8}c) leads to flatter tails of the PDFs;
at the same Prandtl number, increasing the Rayleigh number leads to more extended tails of the PDFs.
These trends generally agree with that in 2D square RB cells \cite{zhang2017statistics} and 3D cylindrical RB cells \cite{emran2008fine}, and can be explained by the positive correlations between increasing Reynolds number and increasing small-scale intermittency of dissipation fields.
To further quantitatively describe the shape of the PDF tails, we adopt a stretched exponential function \cite{chertkov1998intermittent,emran2008fine,zhang2017statistics}
\begin{equation}
p(X^{*}) =\frac{C}{\sqrt{X^{*}}}\exp(-mX^{*\alpha})
\end{equation}
where $C$, $m$ and $\alpha$ are fitting parameters.
$X=\varepsilon_{u,T}/(\varepsilon_{u,T})_{\text{rms}}$ and $X^{*}=X-X_{mp}$ with $X_{mp}$ being the abscissa of the most probable value.
As shown in Fig. \ref{Fig8}, the stretched exponential function (denoted by the solid black lines) fits well with the PDF tails, with adjusted R-squared values above 0.98.

\section{Conclusions}
In this work, we have presented three-dimensional LB simulations of thermal convective flows at high Rayleigh number.
For both laminar thermal convection in side-heated convection cell and turbulent thermal convection in Rayleigh-B\'enard convection cell, the present double distribution function based thermal LB model can give results that agree well with existing benchmark data obtained by other methods.
The extensions to Rayleigh-B\'enard turbulent convection with larger parameter spaces of Rayleigh and Prandtl numbers will be pursued in future work.

\section*{Acknowledgements}
This work was supported by National Natural Science Foundation of China (NSFC) through Grant No. 11772259,
the Fundamental Research Funds for the Central Universities of China (No. 3102019PJ002)
and the 111 project of China (No. B17037).

\section*{References}

\bibliography{mybibfile}

\end{document}